\documentclass[11pt,twoside]{article}

\usepackage{asp2006}
\usepackage{epsf}
\usepackage{lscape}

\newcommand{\pref}{\protect\ref}

\newcommand{\hinode}{{\em Hinode}}

\begin{document}
\title{More of the Inconvenient Truth About Coronal Dimmings}   
\author{Scott W. \textsc{McIntosh}\altaffilmark{1}, Joan \textsc{Burkepile}\altaffilmark{1}, Robert J.  \textsc{Leamon}\altaffilmark{2}}   
\altaffiltext{1}{High Altitude Observatory, National Center for Atmospheric Research, P.O. Box 3000, Boulder, CO 80307, USA}
\altaffiltext{2}{Adnet Systems Inc., NASA Goddard Space Flight Center, Code 671.1, Greenbelt, MD 20771 USA}

\begin{abstract} 
We continue the investigation of a CME-driven coronal dimming from December 14 2006 using unique high resolution imaging of the chromosphere and corona from the Hinode spacecraft. Over the course of the dimming event we observe the dynamic increase of non-thermal line broadening of multiple emission lines as the CME is released and the corona opens; reaching levels seen in coronal holes. As the corona begins to close, refill and brighten, we see a reduction of the non-thermal broadening towards the pre-eruption level. The dynamic evolution of non-thermal broadening is consistent with the expected change of Alfv\'{e}n wave amplitudes in the magnetically open rarefied dimming region, compared to the dense closed corona prior to the CME. The presented data reinforce the belief that coronal dimmings must be temporary sources of the fast solar wind. It is unclear if such a rapid transition in the thermodynamics of the corona to a solar wind state has an effect on the CME itself. 
\end{abstract}

Establishing the poorly understood physical connection between Coronal Mass Ejections (CMEs) and ``transient coronal holes'' \citep[e.g.,][]{Kahler2001} using detailed spectroscopic measurement is a must. Since their initial observation with Skylab \citep[][]{Rust1976} they have come to be viewed as the residual footprint of the CME in the corona \citep[e.g.,][]{Thompson2000}, the radio and plasma signatures of which are observed in interplanetary space \citep[e.g.,][]{Neugebauer1997, Attrill2008}. In this short contributed paper we continue the analysis of \citet{McIntosh2009} (hereafter SWM2009) extending the analysis of \hinode{} \citep{Hinode} Extreme-ultraviolet Imaging Spectrometer \citep[EIS;][]{Culhane2007} data and close by presenting the previously unpublished indicators of CME-forced evolution in the chromosphere and photosphere that are provided by a study of co-teporal Solar Optical Telescope \citep[SOT;][]{Tsuneta2008} data. 

\begin{figure}
\plottwo{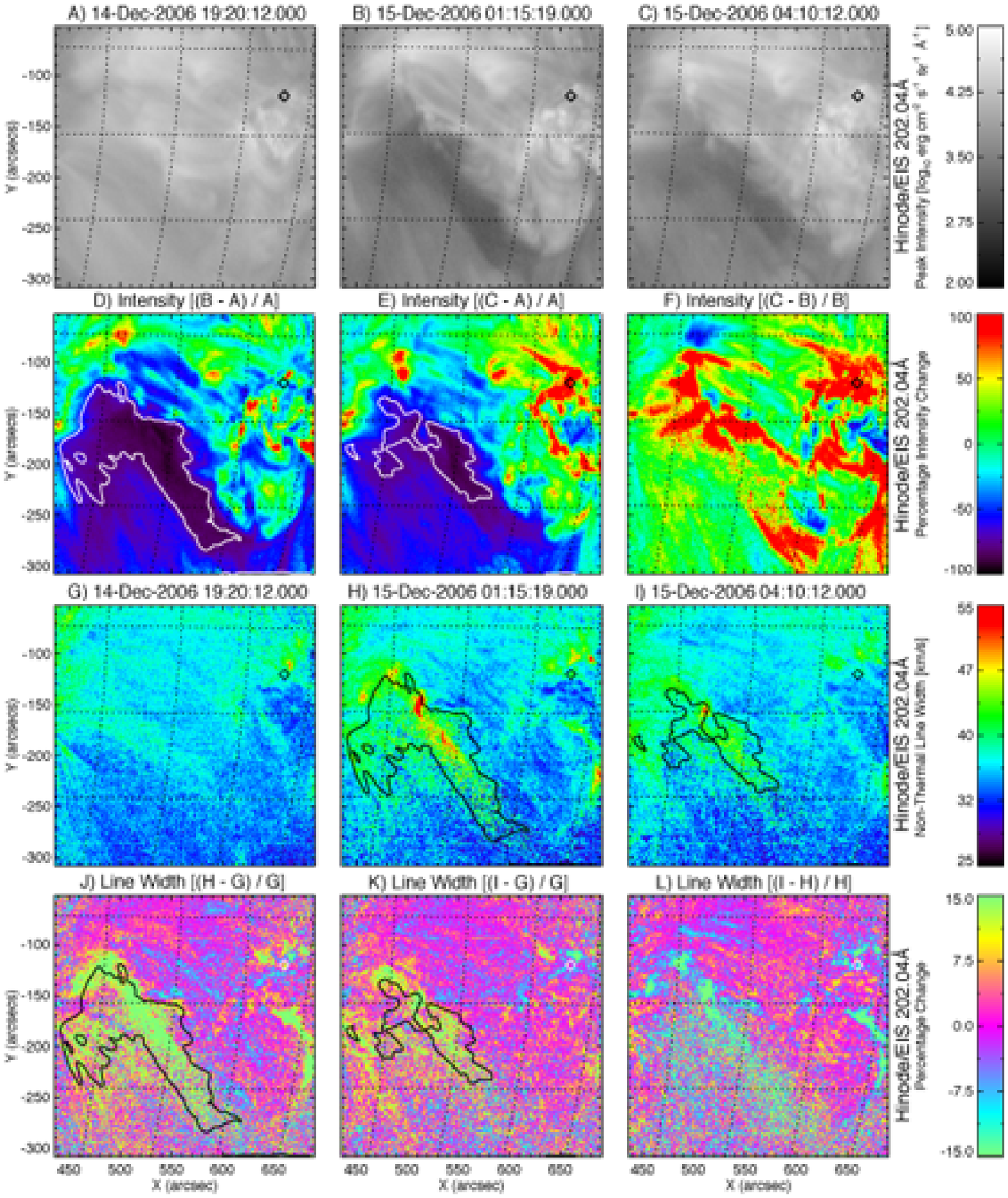}{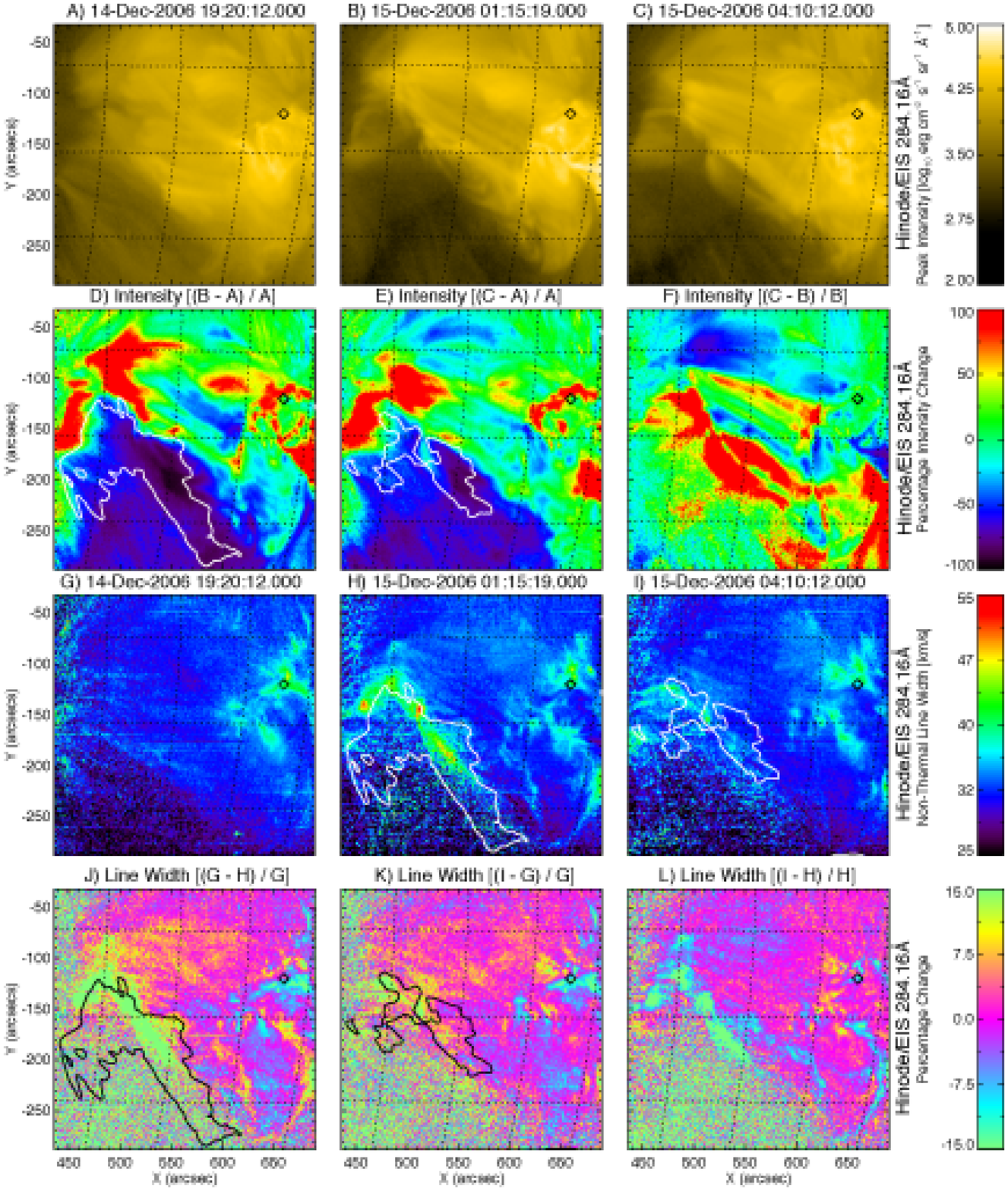}
\caption{Phases of the coronal dimming observed in the intensity and non-thermal line width of the Fe~XIII (left) and Fe~XV (right) emission lines. \label{f1}}
\end{figure}

\begin{figure}
\plotone{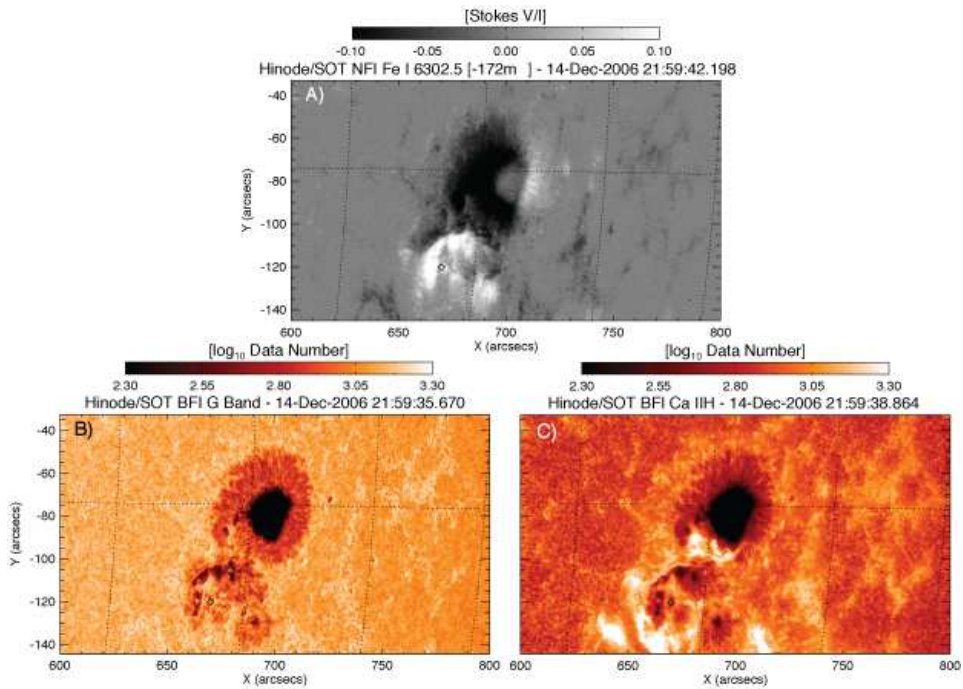}
\caption{SOT observations of Stokes V/I in NFI - Fe I 6302\AA{} (panel A) the BFI G-band (panel B) and Ca~IIH (panel C) channels of the active region nearest to 12/14/2006 22:00UT. Movies are available to accompany the panels of this movie at http://download.hao.ucar.edu/pub/mscott/Hinode2/ \label{f2}}
\end{figure}

\section{Further Observations}   
The EIS dataset comprises of three 256\arcsec $\times$ 256\arcsec spectroheliogram ``raster'' observations (19:20-21:34UT, 01:15-03:30, 04:10-06:24UT) in multiple spectral windows, targeted at the following edge of NOAA~AR~10930. The spectra belonging to the Fe~XIII 202.04\AA{} and Fe~XV 284.16\AA{} lines are fitted and the non-thermal line widths (hereafter $v_{nt}$) are determined using the formalism of SWM2009. In addition, the active region complex was continuously monitored by the focal plane package over the course of the event. The Broadband Filter Imager (BFI) alternately observed Ca~IIH (3968\AA{}) and G-Band (4305\AA{}) from 12/14/2006 19:13 - 12/15/2008 15:58UT with a spatial scale of 0.11\arcsec{} at a cadence ranging from 3~seconds to 5~minutes, but with most frames taken at a 2 minute cadence. The Narrowband Filter Imager (NFI) took shuttered Stokes I \& V measurements at -172m\AA{} from line center of the Fe~I 6302.5\AA{} with a mean cadence of 2~minutes and a spatial scale of 0.16\arcsec{}.

Figure~\pref{f1} shows the three phases of the coronal dimming observed in the intensity and $v_{nt}$ of the Fe~XIII  202.04\AA{} (left) and Fe~XV 284.16\AA{} (right) emission lines. In each case panels A through C show the line intensity before, at the peak and near the end of the dimming event, panels D through F show the percentage change in intensity between panels A and B, C and A and, C and B respectively. Panels G through I show $v_{nt}$ before, at the peak and near the end of the dimming event while panels J through L show the percentage change in $v_{nt}$ between panels H and G, I and G, and, I and H. The solid white contours in panels D and E and the black contours in panels H, I, J and K indicate a 75\% reduction in the Fe~XII 195.12\AA{} line intensity (SWM2009). Like the former analysis we see a large ($\sim$15\%) increase in $v_{nt}$ that is concentrated in and around the perimeter of the 75\% intensity reduction contour. As the dimming event progresses (comparing panels C, A and E), there continue to be regions of difference in $v_{nt}$ late in the lifetime of the dimming, becoming more spatially compact and following the contraction of the intensity decrease contour. In panel F, we see that $v_{nt}$ appears to be slowly recovering to pre-eruption levels.

\subsection{Chicken or Egg? - Evolution in the Deeper Atmosphere}\label{suck}

Close inspection of the movies associated with Fig.~\pref{f2} show that something quite unexpected has happened to the southern portion of the active region during the dimming event. In the region around position 670\arcsec, -120\arcsec (marked with diamonds in all figures) between 22:37UT and 22:44UT the dim penumbra-like emission that we would have commonly associated with cool plasma lying on nearly horizontal magnetic fields completely disappears. This change is most clearly visible in the G-band images, but mirrored in those of Ca~II\footnote{We also notice that by the end of the timeframe considered that a considerable portion of the penumbral structure to the South and East of the region has also gone.}. The penumbral emission disappears as the post-flare/CME brightenings/ribbons move southward over that region. \citet{Gibson2008} indicate that these ribbons are associated with the changing morphology of the magnetic field lines over the active region as the CME lifts off. Interestingly, in panels B and D of Fig.~\pref{f1}, we see that this location shows a significant increase in $v_{nt}$ over the pre-event values that is almost immediately followed by a significant reversal. Unfortunately, there is not enough space in this short article for a detailed discussion of this point or the SOT Spectro-Polarimeter observed the vector field in the vicinity of the active region complex before (12/14/2006 22:00-23:03UT), and after (12/15/2006 05:45-06:48UT) the dimming (see online folder for a graphic of the SP observations). 

Comparing the inverted\footnote{Developed at NCAR under the framework of the Community Spectro-polarimtetric Analysis Center  \-- CSAC; http://www.hao.ucar.edu/projects/csac/} SP observations \citep[][]{Lites2007} of the active region is not straightforward (a considerable amount of time, and subsequent small-scale evolution has occurred between the SP rasters \-- see, e.g., the movie of Fig.~\pref{f2}). There has been a considerable flux cancellation and decrease of field inclination in the region around (680\arcsec, -120\arcsec) leaving a weaker, predominantly positive, polarity mean field. These factors would contribute to the apparent disappearance of the penumbral emission. A significantly more detailed investigation of the small-scale evolution and resulting global magnetic topology changes is required to address this issue. However, we should consider if the change in magnetic topology was instigated by the CME (tied to the dimming) or if it simply related to the gross restructuring of this very complex active region in its effort to find a lower energy state after the eruption \citep[e.g.,][]{Low2001}.

\section{Discussion \& Conclusion}
The spectral information presented reinforces the conclusion of SWM2009 that the changes of the coronal non-thermal line widths are tied to the thermodynamic evolution of the dimming region - and are consistent with the growth and recovery of Alfv\'{e}n wave amplitudes in the varying magnetic topology. Again, we emphasize that, as temporally open magnetic regions, coronal dimmings must be temporary sources of the fast solar wind as the values of $v_{nt}$ reached are equivalent to those observed in coronal holes \citep[][]{McIntosh2009b}. It remains to be seen what impact that fast wind stream, and it's momentum, will have on the CME itself. Clearly, including the rapid thermodynamic variation of the coronal plasma (neglecting changes observed deeper in the atmosphere, Sect.~\pref{suck}) make the understanding of a CME (an intrinsically self-consistent eruption affecting all of the atmospheric layers) much more complex than is currently considered. 

\acknowledgements 
\begin{small}
SWM acknowledges support from NASA grants NNX08AL22G, NNX08AU30G. The National Center for Atmospheric Research is sponsored by the NSF.\end{small}


\begin{thebibliography}{}

\bibitem[{{Attrill} {et~al.}(2008)}]{Attrill2008}
Attrill, G.~D.~R., et~al., 2008, \solphys, 252, 349

\bibitem[{{Culhane} {et~al.}(2007)}]{Culhane2007}
Culhane, J.~L., et~al., 2007, \solphys, 243, 19

\bibitem[{{Gibson} \& {Fan}(2008)}]{Gibson2008}
Gibson, S.~E., Fan, Y., 2008, J. Geophys. Res., 113 (A9), 09103

\bibitem[{{Kahler} \& {Hudson}(2001)}]{Kahler2001}
Kahler, S.~W., \& Hudson, H.~S., 2001, J. Geophys. Res., 106 (A12),  29239

 \bibitem[{{Kosugi} {et~al.}(2007)}]{Hinode}
Kosugi, T., et~al., 2007, \solphys{}, 243, 3

\bibitem[{{Lites} {et~al.}(2007)}]{Lites2007}
Lites, B.~W., Casini, R., Garcia, J., Socas-Navarro, H., Mem. S. A. It., 78, 148

\bibitem[{{Low}(2001)}]{Low2001}
Low, B.~C., 2001, J. Geophys. Res., 106 (A11), 25141

\bibitem[{{Neugebauer} {et~al.}(1997)}]{Neugebauer1997}
Neugebauer, M.,  Goldstein, R., \& Goldstein, B.~E., 1997, J. Geophys. Res., 102, 19743

\bibitem[{{McIntosh}(2009)}]{McIntosh2009}
McIntosh, S.~W., 2009, in press \apj{} Lett. (http://arxiv.org/abs/0809.4024; SWM2009)

\bibitem[{{McIntosh} {et~al.}(2009)}]{McIntosh2009b}
McIntosh, S.~W., Leamon, R.~J., \& De Pontieu, B., 2009, submitted \apj

\bibitem[{{Rust} \& {Hildner}(1976)}]{Rust1976}
Rust, D.~M., Hildner, E., 1976, \solphys, 48, 381

\bibitem[{{Thompson} {et~al.}(2000)}]{Thompson2000}
Thompson, B.~J. et al., 2000, Geophys. Res. Lett., 27, 1431

\bibitem[{{Tsuneta} {et~al.}(2008)}]{Tsuneta2008}
Tsuneta, S., et~al., 2008, \solphys, 249, 167

\end{thebibliography}
\end{document}